\documentclass[12pt]{article}
\usepackage{epsfig}
\usepackage{graphicx}

\usepackage{amsfonts}
 \newcommand{\beq}{\begin{equation}}
 \newcommand{\eeq}{\end{equation}}
 \newcommand{\beqa}{\begin{eqnarray}}
 \newcommand{\eeqa}{\end{eqnarray}}

\newcommand{\erf}{{\rm erf}}
 \def\pabl#1#2{\frac{\partial {#1}}{\partial { #2}}}
 
\begin{document}

\begin{center}

%
%


{\Large{\bf Topological interactions in systems of mutually interlinked polymer rings}}\\[2cm]
{\large{ Matthias Otto}},\\
 Institut f\"ur Theoretische Physik, 

Universit\"at G\"ottingen, Bunsenstrasse 9, D-37073 G\"ottingen,
Germany, \\
otto@theorie.physik.uni-goettingen.de\\[0.5cm]

{\it \today} \\ [2cm]

\end{center}

\underbar{Abstract:}
The
topological interaction arising in interlinked polymeric rings such as
DNA catenanes is considered. More specifically, the free energy for a
pair of linked random walk rings is derived where the distance $R$
between two segments each of which is part of a different ring is kept
constant. The topology conservation is imposed by the Gauss
invariant. A previous approach (M.Otto, T.A. Vilgis,
Phys.Rev.Lett. {\bf 80}, 881 (1998)) to the problem is refined in
several ways. It is confirmed, that asymptotically, i.e. for large $R\gg R_G$ where $R_G$ is
average size of single random walk ring, the effective topological
interaction (free energy) scales $\propto R^4$.


\newpage
\section{Introduction}
Despite their long history, 
topologically constrained system such as knots and links become an
increasingly popular subject of research in statistical physics
\cite{edwards:68,degennes,nechaev:96,kholo:98,grosberg,kardar}.
Chemically synthesized links (called catenanes) have been considered
very early \cite{wasserman:60,frisch:62}. In nature, 
multiply linked rings made of DNA occur in bacteria such as
Escherichia coli during the replication
process as intermediate products \cite{cozzarelli}. Specific topological states (links)
can be formed artificially by turning on and off enzymes, so-called
topoisomerases, which cut and glue the polymeric rings
\cite{cozzarelli} (see also \cite{alberts}).
These methods have been used to study experimentally the
conformational statistics of specific DNA catenanes \cite{levene},
whose elasticity could be studied as well using single-molecule
techniques \cite{arai}.
From the viewpoint of polymer physics topological interactions due to
fixed entanglements remain a challenging
problem for systems such as rubber networks \cite{kremer, everaers}. 


%
\begin{figure}[pb]
  \begin{center} \epsfxsize=10cm
    \epsfig{file=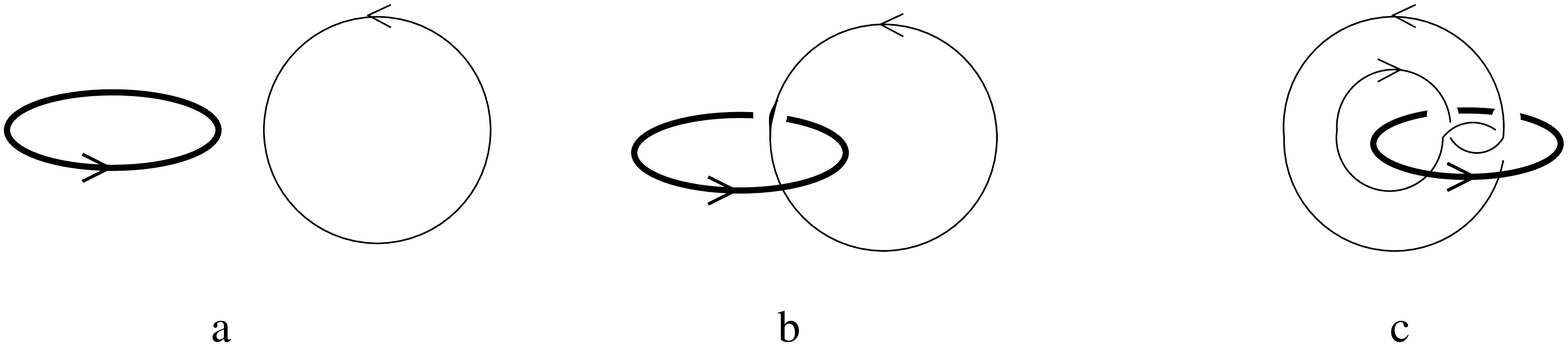,width=10cm,angle=0}
  
  \caption{a) The trivial link: 2 non-entangled rings with linking number $0$. b) Two
    entangled rings with linking number $-1$. c) Whitehead link: two
    entangled rings with linking number $0$. The oriented surface
    bounded by the ``thick'' ring is pierced by the other ring once in
    a positive sense, once in a negative sense.}
  \label{fig1}
\end{center}
\end{figure}

A fundamental problem in theoretical approaches to linked polymer
rings systems is the specification of topological constraints. The
simplest approach introduced by Edwards \cite{edwards:68} is a two rings
invariant, known as the Gauss integral. It is one of the simplest
invariants involving a double line integral of a function based on
respective polymer segment positions of each ring. Thus, it allows to
couple explicitly polymer conformation and topology
conservation. Although being a topological invariant, i.e. being
independent of the specific polymer conformation as long as the
topological state (here the mutual linking number) is conserved, it
has the disadvantage of not being one-to-one. The so-called Whitehead
link (being entangled due to self-intersections one single ring with itself, see figure \ref{fig1})
and the trivial link (consisting of two non-entangled rings) have both the same
linking number zero  (see
\cite{brereton:80} for lucid discussion of this point). This result has roused criticism against the
use of the Gauss invariant in the context of entangled polymer
rings, see e.g. \cite{nechaev:93}. 

The present work responds to these objections in the following way. 
First, we restrict our discussion to random walk
rings which are particularly simple to deal with within an analytical approach. In this case,
a Whitehead link can be disentangled (justifying a linking number zero), because only the
self-entanglement of one rings with itself (see figure \ref{fig1}) keeps the link
from falling apart. The second justification for using the Gauss
integral is the fact that the Gauss integral appears as the first
relevant two-rings invariant appearing within a topological perturbation
series constructed from averages (vacuum-to-vacuum expectation values,
vev's) of Wilson loops with respect to the non-abelian Chern-Simons
theory \cite{guada:93}. The latter has been shown to be equivalent \cite{witten:89} to the HOMFLY knot
polynomial which is an algebraic invariant usually defined in terms of
skein relations, i.e. a recursive equation relating different sets of
crossings (of the 2D projection of a knot or link) giving a recursive algorithm of transforming different knots
and links into each other \cite{guada:93}. In conclusion, the Gauss integral appears
as a quantity to define a minimal model for topology conservation which is well defined
within in a more complete framework just mentioned \cite{otto:diss}.

More specifically, the present work revisits the problem of
calculating the topological interaction for a pair of mutually
interlinked polymer rings. This is accomplished by introducing the constraint that the
distance between two segments each belonging to a different ring be
equal to ${\bf R}$ \cite{otto:98}. Previous work \cite{otto:98}
on the problem
contained several approximations: the distance constraint was not strictly
implemented within conformational averages but
 the distance vector was somehow extracted by a
pre-averaging approximation. This approximation can be given up, as
the present work will show. Moreover the
 discrete nature of the linking number is treated within a systematic
 approach due to Iwata and Kimura \cite{iwata:81} (it has recently received
 some attention in \cite{brereton:01}).
Perhaps not too surprisingly, the anharmonic attraction $\propto R^4$
found in \cite{otto:98} 
 still holds for large distance $R/R_G\gg 1$
where $R_G\sim \sqrt{Nl^2}$ is radius of gyration of a single random
walk ring.


The outline of the paper is as follows. In section 2, we discuss
topology conservation using the Gauss invariant together with a
distance constraint and define conformational averages. In section 3,
the (effective) topological interaction between chain segments
belonging to different rings is calculated: in subsection 3.1, we review
the topological moment expansion due to Iwata and Kimura
\cite{iwata:81}, in 3.2 we derive the topological interaction within
the approximation of using the second topological moment conditional
on a distance constraint $M_2(R)$, before finally giving an explicit
expression for the latter in subsection 3.3. Finally we give results in
subsection 3.4 for the topological interaction and discuss its
dependence on segment length $N$ .
In section 4, a conclusion and a brief outlook is given.

\section{Topology conservation with the Gauss invariant and a distance constraint}
\begin{figure}[hpbt]
  \begin{center} \epsfxsize=10cm
  \epsfig{file=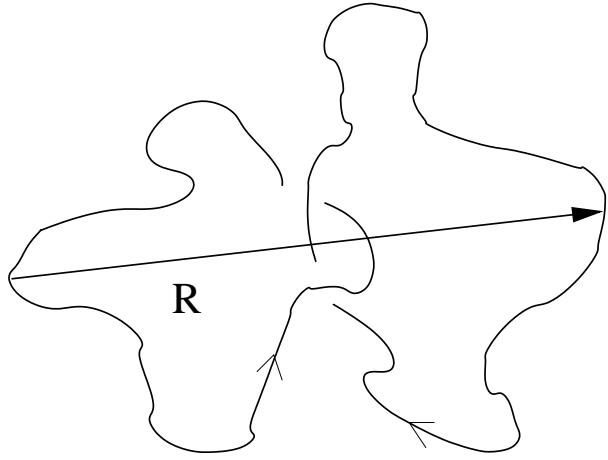,width=8cm,angle=0}
  \end{center}
  \caption{Two entangled rings with two segments located on different rings kept at a distance $R$.}
  \label{fig2}
\end{figure}

We consider a very simple system, i.e. a pair of flexible rings which
are mutually interlinked as in figure \ref{fig2} and which are subject
to an additional distance constraint. The latter controls the distance
between two segments each of which is part of a different ring and is
introduced to monitor the effective topological interaction which arises
due to the topological constraint.

Based on justifications mentiond in the introduction, the topological
constraint imposed on the pair of rings - the fact that their are
wound around each other as in figure \ref{fig2} - is expressed in
terms of the Gauss invariant
\beq
\label{gauss}
\Phi(C_1, C_2)=\Phi_{12}=\frac{1}{4\pi}\oint_{C_1}\oint_{C_2}
d{\bf r}^1\wedge d{\bf r}^2\cdot
\frac{{\bf r}^1-{\bf r}^2}{|{\bf r}^1-{\bf r}^2|^3}
\eeq
The vector functions ${\bf r}^\alpha={\bf r}^\alpha(s)$ for $\alpha=1,2$ denote the
position in three-dimensional space of segments along the polymer
contour, parametrized in terms of $s$, for each ring $C_\alpha$. The
Gauss invariant is always an integer for a given topological state
regardless of the deformations imposed on the polymer conformation,
i.e. the segment position ${\bf r}^\alpha (s)$, which leave the
topological state invariant (siehe z.B.\cite{guada:93}). The integer which is denoted $n$
here is usually called the linking number and counts how often one
chain is wound around the other before being closed to a ring (apart
from a sign which depends on the respective orientation of each ring).

The problem that we propose to study is the partition sum of a given
topological state, i.e. a linking number $n$, for a pair of rings
subject to the distance constraint as shown in figure \ref{fig2}, 
\beq
\label{distance}
Z(n;{\bf R})= \langle\delta(n,\Phi_{12}) \rangle_{\bf R}
\eeq
Inside the average to be defined below a Kronecker delta appears which
is equal to $1$, if the
arguments are equal, and $0$ else.
The average is taken w.r.t. to random walk ring conformations 
\beqa
\label{average}
\langle\dots\rangle_{\bf R}&=&\int{\cal D}{\bf r}^1(s)\int {\cal
  D}{\bf r}^2(s)
\delta({\bf r}^1(0)-{\bf r}^1(N))\delta({\bf r}^2(0)-{\bf r}^2(N))
\nonumber\\
&\times&
\delta({\bf r}^1(0)-{\bf r}^2(0)-{\bf R})
\dots e^{\beta H^1+\beta H^2}\frac{1}{Z_{12,{\bf R}}}
\eeqa
where
\beq
\beta H^\alpha=\frac{3}{2l^2}\int_0^N ds \left(\pabl{{\bf
      r}^\alpha}{s}
\right)^2
\eeq
and $Z_{12,{\bf R}}$ is equal to the numerator on the r.h.s. of
Eq.(\ref{average}) without the dots.
Let us note here that actual averages (carried out in appendix A)
involve a discretized version of the continuous formalism given above.

The problem Eq.(\ref{distance}) has been studied before by Iwata and
Kimura \cite{iwata:81} and more recently by Otto and Vilgis
\cite{otto:98}. In \cite{iwata:81}, the integrand of the Gauss
invariant was however approximated by its behavior close to $|{\bf
  r}^1-{\bf r}^2|\simeq 0$, an approximation which is not
necessary as will be shown below. In \cite{otto:98}, the Gauss
invariant was treated in full generality, but
 a full implementation of the distance
constraint was circumvented by a pre-averaging procedure involving the
tangent vector densities. This approximation can be avoided as will be
shown below. The previously obtained results \cite{otto:98} are in
fact validated by the subsequent analysis.

The choice of the random walk ensemble for polymer conformations is motivated by the search
for an analytical solution. Inclusion of the excluded volume
interactions appears possible in principle (in the context of $O(N)$
field theories \cite{DC}) but seems to render the problem 
of finding an expression for $Z(n;{\bf R})$ rather hopeless.
 A derivation of topological interactions from the Gauss
invariant treating the latter in its most possible generality has not been done
up to now even for the simple case of random walk conformations (except
for a first approach in \cite{otto:98}).

\section{The effective topological interaction}
\subsection{Topological moment expansion}
We next proceed to calculate $Z(n;{\bf R})$. We rewrite the Kronecker
delta inside the brackets in Eq.(\ref{distance}) by introducing a
topological charge, 
\beq
\label{z_r}
Z(n;{\bf R})=\int_{-\pi}^\pi \frac{dg}{2\pi}e^{ign}
\langle 
e^{ig\Phi_{12}}
\rangle_{\bf R}\equiv
\int_{-\pi}^\pi \frac{dg}{2\pi}e^{ign}
\tilde{Z}(g;{\bf R})
\eeq
As already noted in \cite{iwata:81}, the function $\tilde{Z}(g;{\bf
  R})$ can be expanded in terms of topological moments of order $k$, 
$\langle (\Phi_{12})^k \rangle_{\bf R}$. Assuming that $Z(n;{\bf R})$
is an even function of $n$, only the even moments $k=2p$ for $p$ an
integer do contribute. This yields
\beq
\tilde{Z}(g;{\bf R})=\sum_{p=0}^{\infty}\frac{(ig)^{2p}}{(2p)!}M_{2p}({\bf R})
\eeq
where
\beq
M_{2p}({\bf R})=\langle (\Phi_{12})^{2p} \rangle_{\bf R}
\eeq
Next we assume that the $2p$-correlation can be decomposed into
products of $p$ factors involving the correlation $\langle
(\Phi_{12})^{2} \rangle_{\bf R}$ so that
\beq
\label{app1}
\langle (\Phi_{12})^{2p} \rangle_{\bf R}\simeq
\frac{(2p)!}{2^p\,p!}\langle (\Phi_{12})^2 \rangle_{\bf R}^p
\eeq
Then one obtains
\beq
\tilde{Z}(g;{\bf R})=\exp\left(-\frac{g^2}{2}M_2({\bf R})\right)
\eeq
We are thus left with determining the second topological moment given the
distance constraint ${\bf R}$, $M_2({\bf R})$. Before giving the 
explicit functional dependence on ${\bf R}$ , we first derive the topological
interaction for a given linking number and distance constraint in
order to show how $M_2({\bf R})$ appears in the interaction.

\subsection{Topological interaction from the second topological moment}
In order to establish the effective topological interactions between
segments separated by a distance $R=|{\bf R}|$ on different rings
which are mutually entangled, we return to
the evaluation of the constrained partition sum 
$Z(n;{\bf R})$ introduced above.
The second topological moment $M_2(R)$ appears - in the approximation
Eq.(\ref{app1}) given above - inside
the conjugate partition sum $\tilde{Z}(g;{\bf R})$. It is appears
natural now to directly perform the integral in Eq.(\ref{z_r}). However
this leads to negative values for $Z(n;{\bf R})$ for even values of
$n$, while odd values of $n$ give positive values, a problem already
discussed by Iwata and Kimuara\cite{iwata:81}. It is traced back there
to an
unbalanced distribution of errors involved in the approximation of breaking up higher moments
of $\Phi_{12}^2$ as done in Eq.(\ref{app1}). Positive errors are
accumulated for odd $n$ and negative errors for even $n$. \cite{iwata:81}

We therefore proceed like in \cite{iwata:81}, using the ``continuous
Fourier transformation (CFT) method'', whose basic steps are briefly
recalled here.
While $Z(n;{\bf R})$ is defined for discrete $n$, a continuous
function $\zeta(t;{\bf R})$ can be obtained 
by defining
\beq
\zeta(t;{\bf R})=\sum_n \delta(t-n)Z(n;{\bf R})
\eeq
Its Fourier transform reads as
\beq
\tilde{\zeta}(u;{\bf R})=\int_{-\infty}^\infty dt e^{-iut}\zeta(t;{\bf R})
\eeq
On the level of Fourier transforms, $\tilde{\zeta}(u;{\bf R})$ 
and $\tilde{Z}(g;{\bf R})$ are identified for $u=g$ which yields
\beq
\tilde{\zeta}(u;{\bf R})=\tilde{Z}(u;{\bf R})=
\exp\left(-\frac{u^2}{2}M_2({\bf R})\right)
\eeq
The inverse Fourier transform yields
\beqa
\zeta(t;{\bf R})&=&\int_{-\infty}^\infty \frac{du}{2\pi}
e^{iut}\tilde{\zeta}(u;{\bf R})\nonumber\\
&=&\frac{1}{\sqrt{2\pi M_2({\bf R})}}
\exp\left(-\frac{t^2}{2M_2({\bf R})}\right)
\eeqa
Now, the constrained partition sum of discrete linking numbers $n\neq 0$ is
obtained by integrating 
\beq
Z(n;{\bf R})=\int_{n-\frac{1}{2}}^{n+\frac{1}{2}}
dt \zeta(t;{\bf R})
\eeq
In the present case, one finds
\beq
\label{z_top}
Z(n;{\bf R})=\frac{1}{2}
\left(
\erf\left[\frac{|n|+\frac{1}{2}}{\sqrt{2 M_2({\bf R})}}\right]-
\erf\left[\frac{|n|-\frac{1}{2}}{\sqrt{2 M_2({\bf R})}}\right]
\right)
\eeq
where $\erf(x)$ is the error function.
As noted in \cite{iwata:81}, the CFT method leads to the following problem. 
Even though $\zeta(t,{\bf R})$ is assumed to be a multipeak function
in $t$, the final result just obtained is a smooth function in $t$. In
other words the fine structure in $t$ is lost, which is related 
to the approximations imposed on $M_{2p}(R)$ for large $p\gg 1$
\cite{iwata:81}. In the following we will not be interested in the exact
quantitative dependence of $Z(n;{\bf R})$ on $n$, but rather on the
asymptotic behavior for large $R=|{\bf R}|$. Moreover, the errors
contained in the CFT method are evenly distributed among different $n$.

The effective topological interaction can be derived from $Z(n;{\bf
  R})$ as follows
\beq
\label{topint_gen}
\beta F(n;{\bf R})=-\ln Z(n;{\bf R})
\eeq
where $\beta=1/k_B T$ is the inverse thermal energy.
If $M_2({\bf R})=M_2(R)$ is small and/or $n\gg 1$  which happens for large 
$R=|{\bf R}|$ (see below), such that $x_\pm=\frac{|n|\pm\frac{1}{2}}{\sqrt{2 M_2({\bf
      R})}}\gg 1$ 
then the following expansion will be used
\beq
\label{topint}
\beta F(n;{\bf R})=\beta F(n;R)=
\frac{\left(|n|-\frac{1}{2}\right)^2}{2M_2(R)}
+\ln\left[\frac{|n|-\frac{1}{2}}{\sqrt{2\pi M_2(R)}} \right]
+ {\cal O}(1/x_-)
\eeq
We see that the behavior of the topological interaction $\beta F(n;R)$
w.r.t. $R$, in particular for $R\gg l\sqrt{N}$ is dominated by
$1/M_2(R)$.

\subsection{The second topological moment}
We now proceed to calculate $M_2({\bf R})$. 
Using the Fourier
integral 
representation of the Gauss invariant
\beq
\Phi_{12}=-i\int_{\bf q} \int_0^N ds \int_0^N ds'
\dot{r}^1_\mu (s) \dot{r}^2_\nu (s') 
e^{i{\bf q}\cdot({\bf r}^1(s)-{\bf r}^2(s'))}
\epsilon_{\mu\nu\lambda}
\frac{q_\lambda}{q^2}
\eeq
where summation over repeated indices is implied and $\int_{\bf
  q}\dots=\int d^3 q/(2\pi)^3 \dots$,
the second topological moment reads as
\beqa
M_2({\bf R})&=&-\int_{\bf q}\int_{\bf k}
\Pi_{\alpha=1}^4 \left(\int_0^N ds_\alpha\right)
\epsilon_{\mu\nu\lambda}
\epsilon_{\sigma\tau\rho}
\frac{q_\lambda}{q^2}
\frac{k_\rho}{k^2}\nonumber\\
&\times&\left\langle
\dot{r}^1_\mu (s_1) \dot{r}^2_\nu (s_3)
\dot{r}^1_\sigma (s_2) \dot{r}^2_\tau (s_4)
e^{i{\bf q}\cdot({\bf r}^1(s_1)-{\bf r}^2(s_3))}
e^{i{\bf k}\cdot({\bf r}^1(s_2)-{\bf r}^2(s_4))}
\right\rangle_{\bf R}
\eeqa
The term in brackets inside the integrals can be rewritten as
\beqa
\label{corr}
\gamma_{\mu\nu\sigma\tau}&=&
\left\langle
\dot{r}^1_\mu (s_1) \dot{r}^2_\nu (s_3)
\dot{r}^1_\sigma (s_2) \dot{r}^2_\tau (s_4)
e^{i{\bf q}\cdot({\bf r}^1(s_1)-{\bf r}^2(s_3))}
e^{i{\bf k}\cdot({\bf r}^1(s_2)-{\bf r}^2(s_4))}
\right\rangle_{\bf R}
\nonumber\\
&=&
e^{i({\bf q}+{\bf k})\cdot{\bf R}}
F_{\mu\sigma}({\bf q},{\bf k},s_1,s_2)
F_{\mu\tau}(-{\bf q},-{\bf k},s_3,s_4)
\eeqa
The factorization on the r.h.s. is due to the factorization of
conformational averages with respect to polymer $1$ and $2$ in
Eq.(\ref{average}). Details are given in appendix A. We find for
$F_{\mu\sigma}({\bf q},{\bf k},s_1,s_2)$ that
\beqa
\label{F.corr}
F_{\mu\sigma}({\bf q},{\bf k},s_1,s_2) &=&
\frac{1}{2}
\left\{
\frac{l^2}{3}\delta_{\mu\sigma}
\left(\frac{1}{N}-\delta(s_1-s_2)\right)
+\left(\frac{l^2}{3}\right)^2
\left[
\frac{1}{N}(s_1 q_\sigma+s_2 k_\sigma)-k_\sigma
\right]\nonumber\right.\\
&\times&\left.
\left[
\frac{1}{N}(s_1 q_\mu+s_2 k_\mu)-q_\mu-k_\mu
\right]\right\} \nonumber\\
&\times&
\exp\left\{
-\frac{l^2}{6N}
\left[
s_1({\bf q}+{\bf k})^2+(s_2-s_1)k^2-\frac{1}{N}(s_1{\bf q} +s_2{\bf k})^2
\right]
\right\}
\nonumber\\
&+&\left[s_1\leftrightarrow s_2, {\bf q}\leftrightarrow{\bf k}\right]
\eeqa
The last line on the r.h.s. indicates the operations to be done in
order to obtain a fully symmetrized expression (see appendix A). 
Let us note first that the term following the factor
$\left(\frac{l^2}{3}\right)^2$ on the r.h.s. does not contribute to the
second moment, because expressions like  $\dots q_\mu q_\lambda
\epsilon_{\mu\nu\lambda}\dots $ vanish.
In order to further simplify the subsequent analysis, we neglect terms
involving $1/N$ as we assume $N\gg 1$ like in \cite{brereton:92}. Then $F_{\mu\sigma}({\bf
  q},{\bf k},s_1,s_2)$ reads as
\beq
F_{\mu\sigma}({\bf q},{\bf k},s_1,s_2)
\simeq
-\left(\frac{l^2}{3}\right)\delta_{\mu\sigma}\delta(s_1-s_2)
\exp\left\{
-\frac{l^2}{6}s_1 ({\bf q}+{\bf k})^2
\right\}
\eeq
The second topological moment may then be computed as
\beq
M_2({\bf R})=-2\left(\frac{l^2}{3}\right)^2
\int_{\bf q}\int_{\bf k} f({\bf q}+{\bf k},N)^2\frac{{\bf q}\cdot{\bf
    k}}{q^2k^2} e^{i({\bf q}+{\bf k})\cdot{\bf R}}
\eeq
where
\beq
f({\bf u},N)=\int_0^N ds e^{-\frac{l^2}{6}s {\bf u}^2}
=\frac{1}{\left(\frac{l^2}{3}\right){\bf u}^2}\left(1-e^{-\frac{l^2 N}{6} {\bf u}^2}\right)
\eeq
Transforming ${\bf k}\rightarrow{\bf u}={\bf q}+{\bf k}$, one obtains
\beq
M_2({\bf R})=-8
\int_{\bf q}\int_{\bf u} \frac{{\bf q}\cdot{({\bf u}-{\bf
      q})}}{q^2|{\bf u}-{\bf q}|^2} \frac{e^{i{\bf u}\cdot{\bf R}}}{u^4}
\left(1-e^{-\frac{l^2 N}{6}u^2}\right)^2
\eeq
Introducing ${\bf u}'=\sqrt{l^2N/(6)}{\bf u}$ and ${\bf X}={\bf
  R}/\sqrt{l^2N/(6)}$ and dropping primes, the double integral is
slightly simplified:
\beq
\label{m2.1}
M_2({\bf R})=-8
\int_{\bf q}\int_{\bf u} \frac{{\bf q}\cdot{({\bf u}-{\bf
      q})}}{q^2|{\bf u}-{\bf q}|^2} \frac{e^{i{\bf u}\cdot{\bf X}}}{u^4}
\left(1-e^{-u^2}\right)^2
\eeq
The integration which involves some approximations affecting the short
distance behavior (to be discussed below) and which is given in
appendix B, yields
\beq
\label{2top}
M_2(R)=
\frac{1}{8\pi^3}
\left(
\Lambda\frac{\sqrt{\pi}}{2}
-\pi\,{}_1 F_1\left(-\frac{1}{2};\frac{3}{2};\frac{X^2}{4}\right)
\right)e^{-X^2/4}
\eeq
where $R=X\sqrt{l^2N/6}$ and ${}_1 F_1\left(\alpha;\beta;x\right)$ is
the confluent hypergeometric function. The expression involves a cutoff
$\Lambda\sim \sqrt{N}$ (see appendix A) whose relevance will be
discussed below.
It is interesting to note what happens when $M_2(R)$ is integrated
over all distances between the segments. Then the unconditional second
topological moment is obtained as
\beqa
M_2&=&V^{-1}\int d^3 R M_2(R)\nonumber\\
&=&V^{-1}\sqrt{l^2N/6}^3 4\pi\int_0^\infty dX X^2 
\frac{1}{8\pi^3}
\left(
\Lambda\frac{\sqrt{\pi}}{2}
-\pi \,{}_1 F_1\left(-\frac{1}{2};\frac{3}{2};\frac{X^2}{4}\right)
\right)e^{-X^2/4}\nonumber\\
&\sim& V^{-1} l^3 N^{3/2}
\left(c_1 N^{1/2}-c_2
\right)
\eeqa
where in the last line $\Lambda\sim\sqrt{N}$ has been used and 
where $c_1$ and $c_2$ are numerical factors. Note that the factor
$1/V$ on the r.h.s. of
the previous equation is understood in the limit $V\rightarrow\infty$ 
and $N\rightarrow\infty$ such that $\rho = N/V$ is a density. This
situation is given if
 2 rings fill out a macrocopic volume. Then the unconditional second 
topological moment has the scaling 
\beq
M_2\sim \rho l^3\left( c_1 N - c_2 N^{1/2}\right)
\eeq
a result,
which up to numerical factors $c_1$ and $c_2$ has been found for the topological moment of randomly
entangled polymer rings in a dense system where the excluded volume
effect is ignored \cite{ferrari:00, otto_bf}. 
Therefore, the cutoff-dependent first term on the r.h.s. of Eq.(\ref{2top})
appears essential in order to establish consistency with the
well known result for $M_2$ for random walk rings to scale to dominant
order like $\propto N$.
We will see below that it affects the effective segment-segment
interaction only for small up to intermediate deformations of the rings.

As shown in the previous section, the behavior of $1/M_2(R)$ is particularly important to obtain 
the large $R$ behavior of the topological interaction.
Noting that 
\beq
{}_1 F_1\left(-\frac{1}{2};\frac{3}{2};\frac{X^2}{4}\right)
=e^{X^2/4}{}_1 F_1\left(2;\frac{3}{2};-\frac{X^2}{4}\right)
\eeq
and using for large $X\gg 1$
\beq
{}_1 F_1\left(2;\frac{3}{2};-\frac{X^2}{4}\right)
\simeq -\frac{4}{X^4}\left(1+\frac{12}{X^2}+\dots\right)
\eeq
one obtains
\beqa
\frac{1}{M_2(R)}&\simeq& 
8\pi^3\left(\frac{\sqrt{\pi}}{2}\Lambda\exp(-X^2/4)+\frac{4\pi}{X^4}\right)^{-1}\nonumber\\
&\simeq& 2\pi^2
X^4
\eeqa
for large $X=R/\sqrt{Nl^2/6}$.
It gives the following effective topological interaction 
\beq
\label{topint_large}
\beta F(n;X)\simeq
\pi^2\left(|n|-\frac{1}{2}\right)^2 X^4
\eeq
Concerning the $X^4$ dependence, this result agrees perfectly with the
previously derived asymptotics for the topological interaction of two
concatenated chains
based on a pre-averaging approximation \cite{otto:98}.

\subsection{Results}
\begin{figure}[hbt]
  \begin{center} 
  \epsfig{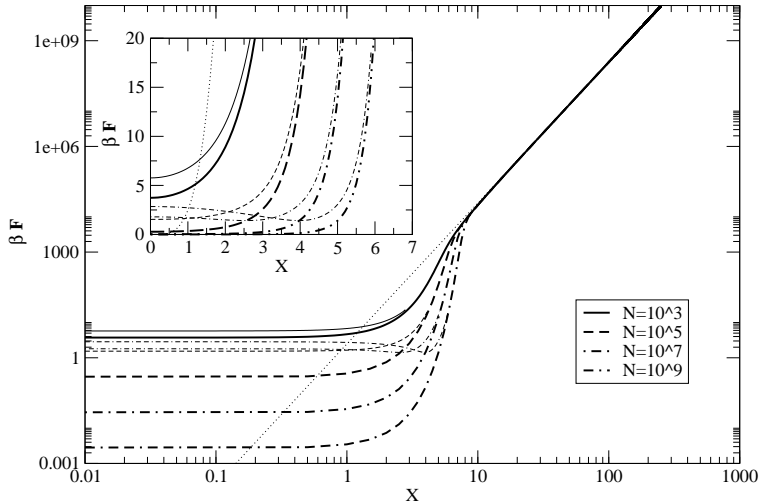}
  \end{center}
  \caption{The topological interactions for linking number $n=1$ based
    on, one,  the first term in
    Eq.(\ref{topint}) (thick lines), and
    two, on the full expression Eq.(\ref{z_top}) for the partition
    function (thin lines), both together with Eq.(\ref{2top}). The thin dotted line corresponds to the
    asymptotic expression Eq.(\ref{topint_large}). The inset shows in
    a double-linear plot how the curves based on the full expression
    Eq.(\ref{z_top}) approach the approximate ones (Eq.(\ref{topint})).}
  \label{fig3}
\end{figure}

Given the expression for the conditional topological moment
Eq.(\ref{2top}), we consider now the effective topological interaction
in more detail. 
For large $X$, we use the 
 first term in the expansion 
Eq.(\ref{topint}). For small $X$ where the Eq.(\ref{topint}) is bound
to fail, the exact expression Eq.(\ref{z_top}) for the partition
function (based on the second topological moment conditional on $R$ or
$X$) is used. 
In Fig.  \ref{fig3}, curves for small $X$ (obtained numerically from Eq.(\ref{z_top})) for
the topological interaction $\beta F$ for linking number $n=1$ for various chain lengths
$N$ are shown, together with the corresponding curves for all $X$
calculated from the first term in Eq.(\ref{topint}), whith both sets
of curves involving Eq.(\ref{2top}) for the 2nd topological moment $M_2(R)$.
\vspace{2cm}

The topological interactions computed from the full
expression Eq.(\ref{z_top}) smoothly approach 
the approximate ones (valid for large $X$)
 based on 
Eq.(\ref{topint}). 
Whereas for large $X$ the topological interaction is independent of
the chain length $N$ approaching the asymptotic form Eq.(\ref{topint_large}), such a dependence appears for intermediate and
small $X$. To be precise, the interaction 
 becomes weaker for larger $N$ for a given
intermediate $X$, just below $X=10$. This effect might be due to the discrete polymer model
used here. Let us consider two systems of two concatenated rings which
are similar in their conformation but which have different chain
length $N_1$ and $N_2$ with $N_2>N_1$. Then the system with chain
length $N_2$
suffers a smaller entropy loss due to the topological constraint. Thus
the topological interaction is reduced.

However, for very small $X$ the topological interaction does not show
a monotonuous dependence on $N$ but passes through a minimum as $N$ is
increased from $10^3$ to $10^9$.
Moreover, for very large $N$ a minimum of the topological interaction
as function of $X$
starts to develop. 
A strong repulsion of segments belonging to different rings at small
$X$ found in \cite{otto:98} is not confirmed by the present
calculation. On the other hand, the present work depends on
approximations (see appendix B) which possibly influence the short distance
behavior of the topological interaction. Therefore, no conclusive
statement is claimed for the topological interactions derived from the
Gauss invariant on small scales $X$. A more complete analysis of the
short-distance behavior is necessary and the subject of future work.

For extreme elongations $X\gg 1$, one might expect for the topological
interaction a crossover from $X^4$ to $X^2$ as claimed in \cite{brereton:01}, based on the argument
that when two random walk rings 
linked with linking number $1$ are fully stretched, they form an
effective random walk chain. Apparently, this limit cannot be reached
by the present approach. The argument just given seems to  assume,
however,  some local
gluing and some partial inextensibility of the chains which then leads
to a mutual attachment of the rings due to the topological constraint.
Each ring conformation is however treated here w.r.t. to the ensemble of
random walks. Therefore the effective conformation of a single random
walk chain is not reached.

\section{Discussion}
In the present work, the topological interaction between segments 
belonging to different polymer rings which are mutually interlinked is
considered. Two segments are singled out and are kept at a distance
$R$, while the conformations of each ring is averaged over w.r.t. a
random walk ensemble. Supplementing
a previous calculation \cite{otto:98}, the present analysis introduces 
several refinements. First,
the distance constraint is implemented from the very beginning. The discrete nature of the linking number is
treated systematically following \cite{iwata:81}.

The topological interaction behaves asymptotically, i.e. for $X\gg 1$
where $X=R/\sqrt{Nl^2/6}$,  as $\sim X^4$ as found previously
\cite{otto:98}. For smaller $X\leq 10$, the topological interaction
depends on the chain length $N$. For intermediate distances $X$, an
argument related to the reduced entropy loss due to the topology
conservation for increasing $N$ can be given. For small $X$, no
final conclusive statement on the topological interaction can be
given, as the approximations used for the conformational averages in
order to obtain analytical results possibly affect small length
scales. We remark however that the present result is consistent with
calculations of the unconditional second topological moment \cite{ferrari:00, otto_bf}.

The present work may be considered as a further step forward to understand the
elasticity of
DNA catenanes (interlinked rings) which can be studied in principle
using single-molecule techniques \cite{arai}. A first step was the
previous study \cite{otto:98} in order to understand the
conformational statistics of DNA catenanes \cite{levene}. The simple random walk
model used here is of course not realistic in this
context. Apart from the excluded volume problem, DNA
molecules are semiflexible chains \cite{marko} which requires the Kratky-Porod
model \cite{KP} which however leads to analytical problems. Nonetheless, for
long 
chains, self-avoiding walk behavior is recovered.

Possible future directions of the problem presented here not only concern
to use more realistic conformational models relevant to the study of
biomolecules. Future work should aim at a more accurate treatment of
conformational averages, in particular the inclusion of terms which
were omitted here for large $N$ (which amount to neglect the closure
of chains). Another important problem is the
issue of localization: i.e. the hypothesis, that a number of segments
are localized close to a region containing the link, motivating
concepts such as slip-links (see e.g.\cite{edwards:88}) which have found renewed
interest \cite{metzler:02}.\\

\noindent
{\bf Acknowledgements}:
Financial support by the DFG under
grant Zi209/6-1 is gratefully acknowledged.

\section*{Appendix A: The correlation function
  $\gamma_{\mu\nu\sigma\tau}$ }
Following \cite{brereton:92}, 
we calculate the following correlation function of Eq.(\ref{corr})
\beq
\gamma_{\mu\nu\sigma\tau}=
\left\langle
\dot{r}^1_\mu (s_1) \dot{r}^2_\nu (s_3)
\dot{r}^1_\sigma (s_2) \dot{r}^2_\tau (s_4)
e^{i{\bf q}\cdot({\bf r}^1(s_1)-{\bf r}^2(s_3))}
e^{i{\bf k}\cdot({\bf r}^1(s_2)-{\bf r}^2(s_4))}
\right\rangle_{\bf R}
\eeq
To do so, we introduce a discretized random walk ${\bf r}^1(s)={\bf
  R}^1_0+\sum_{i=1}^s {\bf b}^1_i$ such that the measure of the path
integral in Eq.(\ref{average}) becomes
\beq
{\cal D}{\bf r}^1(s)\rightarrow \int d^3 R^1_0 \prod_{i=1}^N \int d^3 b^1_i
\eeq
Then the correlation function given above is derived from the
following generating functional
\beq
G\left(\{{\bf h}^1_i\}, \{ {\bf h}^2_i\}; {\bf q}, {\bf k}; s_1, s_2, s_3,
  s_4\right)
=\left\langle
e^{i{\bf q}\cdot({\bf r}^1(s_1)-{\bf r}^2(s_3))}
e^{i{\bf k}\cdot({\bf r}^1(s_2)-{\bf r}^2(s_4))}
\right\rangle_{\{ {\bf R}\}, \{{\bf h}^1_i\}, \{ {\bf h}^2_i\}}
\eeq
The average is understood  as
\beqa
&&G\left(\{{\bf h}^1_i\}, \{ {\bf h}^2_i\}; {\bf q}, {\bf k}; s_1, s_2, s_3,
  s_4\right)=
\frac{1}{Z_{12,{\bf R}}}
\int d^3 R^1_0 \prod_{i} \int d^3 b^1_i
\int d^3 R^2_0 \prod_{i} \int d^3 b^2_i
\nonumber\\
&\times&
\delta(\sum_i {\bf b}^1_i)\delta(\sum_i {\bf b}^2_i)
\delta({\bf R}^1_0-{\bf R}^2_0-{\bf R})
\exp\left[-\beta H^1-\beta H^2+\sum_{\alpha=1,2} \sum_i  {\bf
    h}^\alpha_i \cdot {\bf b}^\alpha_i \right]
\nonumber\\
&\times&
\exp\left[
-i({\bf q}+{\bf k})\cdot {\bf R}^1_0+i({\bf q}+{\bf k})\cdot {\bf R}^2_0
\right]
\nonumber\\
&\times&
\exp\left[
-i{\bf q}\cdot\sum_{i=1}^{s_1}{\bf b}^1_i-i{\bf k}\cdot\sum_{i=1}^{s_2}{\bf b}^1_i+i{\bf q}\cdot\sum_{i=1}^{s_3}{\bf b}^2_i+i{\bf k}\cdot\sum_{i=1}^{s_4}{\bf b}^2_i
\right]
\eeqa
Products and sums in the equation above marked by an index $i$ only
are supposed to cover the range $i=1..N$.
The integrations w.r.t. to ${\bf R}^\alpha_0$ for $\alpha=1,2$ in the
numerator and denominator on the r.h.s. of the equation above are
easily carried out to give
\beqa
\label{split}
G\left(\{{\bf h}^1_i\}, \{ {\bf h}^2_i\}; {\bf q}, {\bf k}; s_1, s_2, s_3,
  s_4\right)&=&e^{i({\bf q}+{\bf k})\cdot {\bf R}}
F\left(\{{\bf h}^1_i\},
{\bf q},{\bf k}, s_1, s_2
\right)\nonumber\\
&\times&
F\left(\{{\bf h}^2_i\},
-{\bf q},-{\bf k}, s_3, s_4
\right)
\eeqa
where
\beqa
\label{F}
F\left(\{{\bf h}_i\},
{\bf q},{\bf k}, s_1, s_2
\right)&=&
\frac{1}{Z_{1}}
 \prod_{i} \int d^3 b_i
\delta(\sum_i {\bf b}_i)
\exp\left[-\frac{3}{2l^2}\sum_i ({\bf b}_i)^2+ \sum_i  {\bf
    h}_i \cdot {\bf b}_i \right]
\nonumber\\
&\times&
\exp\left[
-i{\bf q}\cdot\sum_{i=1}^{s_1}{\bf b}_i-i{\bf k}\cdot\sum_{i=1}^{s_2}{\bf b}_i
\right]
\eeqa
The partition sum $Z_1$ is the single ring partition function devided
by a volume factor (originating from the integrations w.r.t. to ${\bf
  R}^\alpha_0$ for $\alpha=1,2$ which have already been carried out).
Now one has to choose an order for $s_1$, $s_2$ which are now
considered as being discrete out of $1..N$, say $s_1\leq s_2$. Then
the integrations are easily carried out to give
\beqa
\label{F.result}
F\left(\{{\bf h}_i\},
{\bf q},{\bf k}, s_1, s_2
\right)&=&
\exp\left[
\frac{l^2}{6N}\left(
\sum_i {\bf h}_i +s_1 {\bf q}+ s_2{\bf k}
\right)^2
\right.\nonumber\\
&-&\left.\frac{l^2}{6}
\left(
\sum_{i=1}^{s_1}\left({\bf h}_i +{\bf q}+ {\bf k}\right)^2
+\sum_{i=s_1+1}^{s_2}\left({\bf h}_i +{\bf k}\right)^2
+\sum_{i=s_2+1}^{N}{\bf h}_i^2
\right)
\right]\nonumber\\
\eeqa
The case $s_1>s_2$ is obtained by interchanging $s_1\leftrightarrow
s_2$ and ${\bf q}\leftrightarrow {\bf k}$.
Now, the correlation function $\gamma_{\mu\nu\sigma\tau}$ is obtained
as follows
\beq
\gamma_{\mu\nu\sigma\tau}=
\frac{1}{(-i)^4}
\pabl{}{h^1_{\mu,s_1}}
\pabl{}{h^2_{\nu,s_3}}
\pabl{}{h^1_{\sigma,s_2}}
\pabl{}{h^2_{\tau,s_4}}
G\left(\{{\bf h}^1_i\}, \{ {\bf h}^2_i\}; {\bf q}, {\bf k}; s_1, s_2, s_3,
  s_4\right)|_{{\bf h}^1_i={\bf 0}, {\bf h}^2_i={\bf 0}}
\eeq
Using Eq.(\ref{split}) together with Eq.(\ref{F.result}) one obtains
Eq.(\ref{corr}) with Eq.(\ref{F.corr}) for $\gamma_{\mu\nu\sigma\tau}$ given in the main text.

\section*{Appendix B: Approximate evaluation of integrals}
Defining the integral w.r.t. ${\bf q}$ in Eq.(\ref{m2.1}) as $l({\bf u})$ one obtains
\beq
M_2({\bf R})=-\frac{16}{(2\pi)^2}\frac{1}{X}
\int_0^\infty du\frac{\sin(uX)}{u^3}\left(1-e^{-u^2}\right)^2 l(u)
\eeq
where 
\beq
l(u)=l({\bf u})=\int_{\bf q}\frac{{\bf q}\cdot{({\bf u}-{\bf
      q})}}{q^2|{\bf u}-{\bf q}|^2}
=\frac{2}{(2\pi)^2}B(3/2,3/2)\int_0^\Lambda dq \frac{u^2q^2-q^4}{(q^2+u^2)^2}
\eeq
Note that the integral in the last eq. needs to be cut off by a
parameter $\Lambda$ which is related to the cutoff in momentum space 
$\Lambda'$ via $\Lambda=\Lambda' \sqrt{l^2N/6}$ and thus 
proportional to $\sqrt{N}$ if $\Lambda'\sim l^{-1}$. Evaluating
the $q$-integral yields for $l(u)$
\beq
l(u)=\frac{1}{(2\pi)^2}\frac{\pi}{8}
\left(\frac{\pi}{2}u-2\Lambda +3u\arctan(\Lambda/u)-\frac{\Lambda u^2}{\Lambda^2+u^2}
\right)
\eeq
In order to make analytical progress, let us approximate
$\left(1-e^{-u^2}\right)^2/u^3\simeq u e^{-u^2}$ in Eq.(\ref{m2.1}). The error involved
in this approximation concerns large $u$, and the short distance
behavior of $M_2$, which is discussed in the main text. Then the second
topological moment is given by the expression
\beq
M_2(R)=-\frac{16}{(2\pi)^2}\frac{1}{X}
\int_0^\infty du\, l(u) u \sin(uX) e^{-u^2}
\eeq
Inside the integral $l(u)$ is approximated, for $\Lambda\gg 1$, as
\beq
\label{l.app}
l(u)\simeq\frac{1}{(2\pi)^2}\frac{\pi}{8}
\left(-2\Lambda+ 2\pi u-\frac{u^2}{\Lambda}
\right)
\eeq
where $\arctan(\Lambda/u)\simeq \pi/2$ for $\Lambda\gg 1$.
Using the approximation above, evaluation of the integral w.r.t. $u$
involves the following integrals. 
The term of order 0 in $u$ gives the integral
\beq
I_1=\int_0^\infty du\, u \sin(uX) e^{-u^2}
=\frac{\sqrt{\pi}X}{4}e^{-X^2/4}
\eeq
The term linear in
$u$ in $l(u)$ yields the integral
\beq
I_2=\int_0^\infty du\, u^2 \sin(uX) e^{-u^2}
=\frac{X}{2}e^{-X^2/4}  \,{}_1 F_1\left(-\frac{1}{2};\frac{3}{2};\frac{X^2}{4}\right)
\eeq
while the
term of order 2 in $u$ involves evaluating
\beq
I_3=\int_0^\infty du\, u^3 \sin(uX) e^{-u^2}
=\frac{X}{2}e^{-X^2/4}\Gamma(5/2)\, {}_1 F_1\left(-1;\frac{3}{2};\frac{X^2}{4}\right)
\eeq
The function ${}_1 F_1\left(-1;\frac{3}{2};\frac{X^2}{4}\right)
$ explicitly gives $1-X^2/6$. However, we will neglect the $u^2$ term
in $l(u)$ because it is multiplied by a factor $1/\Lambda$ (see Eq.(\ref{l.app})). Collecting terms, one obtains for the second moment
\beq
\label{2top.app}
M_2(R)=
\frac{1}{8\pi^3}
\left(
\Lambda\frac{\sqrt{\pi}}{2}
-\pi\,{}_1 F_1\left(-\frac{1}{2};\frac{3}{2};\frac{X^2}{4}\right)
\right)e^{-X^2/4}
\eeq
where $R=X\sqrt{l^2N/6}$ and ${}_1 F_1\left(\alpha;\beta;x\right)$ is
the confluent hypergeometric function (see Eq.(\ref{2top}) in the main
text).

\end{document}